\documentclass[prb,twocolumn, floatfix, superscriptaddress]{revtex4} 
\usepackage{graphicx}
\usepackage{epstopdf}
\usepackage{bm, amsmath, amssymb}

\newcommand{\nbar}[0]{\bar{n}}

\newcommand{\adag}[0]{a^\dagger}

\begin{document}

\title{Cavity-assisted quantum bath engineering}

\author{K. W. Murch*}
\affiliation{Quantum Nanoelectronics Laboratory, Department of Physics, University of California, Berkeley CA 94720}

\author{U. Vool}
\affiliation{Department of  Physics, Yale University, New Haven, CT 06520-8120}
\author{D. Zhou}
\affiliation{Department of  Physics, Yale University, New Haven, CT 06520-8120}
\author{S. J. Weber}
\affiliation{Quantum Nanoelectronics Laboratory, Department of Physics, University of California, Berkeley CA 94720}
\author{S.M. Girvin}
\affiliation{Department of  Physics, Yale University, New Haven, CT 06520-8120}
\author{I. Siddiqi}
\affiliation{Quantum Nanoelectronics Laboratory, Department of Physics, University of California, Berkeley CA 94720}

\date{\today}

\begin{abstract}
We demonstrate quantum bath engineering for a superconducting artificial atom coupled to a microwave cavity.  By tailoring the spectrum of microwave photon shot noise in the cavity, we create a dissipative environment that autonomously relaxes the atom to an arbitrarily specified coherent superposition of the ground and excited states.  In the presence of background thermal excitations, this mechanism increases the state purity and effectively cools the dressed atom state to a low temperature.
 \end{abstract}

\maketitle

 In practice, quantum systems are never completely isolated, but instead interact with degrees of freedom in the surrounding environment,  eventually leading to decoherence of some states of the system.  Precision measurement techniques such as nuclear magnetic resonance and interferometry,  as well as envisioned quantum schemes for computation, simulation, and data encryption, rely on the ability to prepare and preserve delicate quantum superpositions and entanglement.  The conventional route to long-lived quantum coherence involves minimizing coupling to a dissipative bath.  Paradoxically, it is possible to instead engineer specific couplings to a quantum environment that allow dissipation to actually preserve coherence\cite{poya96qre,krau08,krau11,marcos2012}.
In this letter, we demonstrate such quantum bath engineering for a superconducting artificial atom coupled to a microwave frequency cavity.  Cavity-assisted cooling of the atom is tailored to produce any arbitrary superposition of ground and excited states on demand with high fidelity.

The concept of our experiment is shown in Figure 1.
A two-level atom is driven resonantly at frequency $\omega_\mathrm{q}$. In the frame rotating at the drive frequency, the two eigenstates of the system are $|\pm\rangle =( |g\rangle \pm |e\rangle)/\sqrt{2}$, with eigenvalues $\pm1$ 
of the $\sigma_x$ Pauli operator.   The energy splitting between the $|+\rangle$ and $|-\rangle$ states is given by the Rabi frequency
, $\Omega_\mathrm{R}$.  If $\hbar \Omega_\mathrm{R}\ll k_\mathrm{B} T_\mathrm{eff}$, where $T_\mathrm{eff}$ is the effective temperature, neither state is thermodynamically preferred.  However, by weakly coupling the atom to a cavity and introducing an additional drive detuned from the cavity resonance by $\Delta_\mathrm{c} = \omega_\mathrm{d} - \omega_\mathrm{c}$, the photon shot noise of the cavity forms a quantum bath for the atom which can be engineered such that dissipation drives the atom to the $|+\rangle$ or $|-\rangle$ state. Here, $\omega_\mathrm{d}(\omega_\mathrm{c})$ is the drive (cavity resonance) frequency.  For red detuned drive, $(\Delta_\mathrm{c}<0$), the cavity dissipation ``cools'' the atom to the $|+\rangle$ state.  

Cavity assisted cooling has been studied extensively in the context of atomic gases\cite{hora97cavity,vule00cool,maun04cool,leib09sideband}, mechanical objects\cite{arci06,giga06cooling,naik06back,schl06cooling} and spins \cite{brah10spin}. Similarly, single atoms and qubits have been used to alter the dissipation environment of a resonator, leading to lasing\cite{mckeever03laser,haus08lasing}, cooling, and amplification\cite{graj03coolandamp,PhysRevLett.91.097906,oels12}.   We demonstrate that
cavity cooling can be applied to the dressed states of a two-level atom, and the dissipation
introduced by the drive
may be engineered to relax the system towards any specified point on the Bloch sphere\textemdash a valuable resource in quantum information processing. This process is resonant and can produce large cooling rates.

\begin{figure} [htp]
\includegraphics[angle = 0, width = 0.5\textwidth]{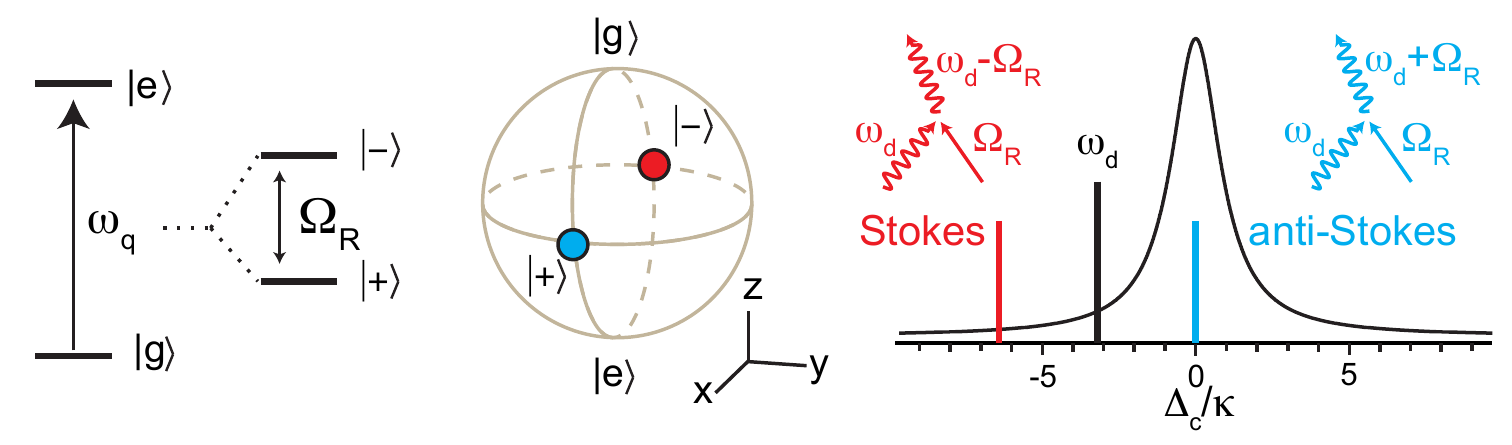}
\caption{\label{fig1}  Cavity cooling of a dressed state.   A two-level atom is driven resonantly to form new eigenstates, $|+\rangle$ and $|-\rangle$ in the rotating frame.  Driving the cavity at $\Delta_\mathrm{c} = -\Omega_\mathrm{R}$ resonantly enhances the anti-Stokes process relaxing the system to the $|+\rangle$ state. 
}
\end{figure}

Our two-level atom is realized using the two lowest energy levels of a  superconducting transmon qubit\cite{koch07transmon, paik113D} with $\omega_\mathrm{q}/2\pi = 5.0258$ GHz,   dispersively coupled to the TE$_{101}$ mode of a 3D superconducting cavity with frequency $\omega_\mathrm{c}/2 \pi =6.826$ GHz and linewidth $\kappa /2\pi = 4.3$ MHz.  The qubit induces a state-dependent frequency shift on the cavity of $-\chi\sigma_z$ where $\chi/2\pi = -0.66$ 
MHz is the dispersive coupling strength.  Similarly, the qubit frequency undergoes a light (or AC Stark) shift depending on the intracavity photon number, $\hat n$, with mean value denoted $\omega_\mathrm{q}' = \omega_\mathrm{q}+2\chi \nbar$.
The Hamiltonian for the qubit in the frame rotating at $\omega_\mathrm{q}$ is,
\begin{equation}
H = -\frac{\Omega_\mathrm{R}}{2} \sigma_x - \chi a^\dagger a \sigma_z, 
\end{equation}
where $\adag(a)$ is the cavity photon creation (annihilation) operator.
Transition rates between the $|+\rangle$ and $|-\rangle$ states are determined from Redfield theory\cite{cler10noise}, $\Gamma_\pm = [4\chi^2S_{nn}(\mp\Omega_\mathrm{R}) + \tilde{S}_{yy}(\mp\Omega_\mathrm{R})+\tilde{S}'_{zz}(\mp\Omega_\mathrm{R})]/4$, where $\tilde{S}_{yy} \simeq 1/(T_1)$ and $\tilde{S}'_{zz} \simeq2/( T_\varphi)$ are the power spectral densities of noise orthogonal to the $x$ axis in the rotating frame, and $S_{nn}(\omega) = \nbar \kappa[(\kappa/2)^2+(\omega+\Delta_\mathrm{c})^2]^{-1}$ is the spectral density of photon number fluctuations in the cavity\cite{cler10noise} that characterize the quantum bath. 
$T_1$ and $T_\varphi$ are the energy decay and pure dephasing times for the qubit, respectively. For $\Delta_\mathrm{c} = 0$, $S_{nn}$ is symmetric in frequency and corresponds to an infinite temperature bath.  In this case $S_{nn}$ causes dephasing of the qubit  and can equivalently be described in terms of a fluctuating AC stark shift, or in terms of measurement induced dephasing\cite{cler10noise,bois09}. When $\Delta_\mathrm{c}\neq0$, $S_{nn}$ is asymmetric and corresponds to a bath with finite positive (or negative) temperature and can be used to cool (or invert) the qubit state\cite{cler10noise}. As illustrated in Figure 1, cooling takes place via inelastic Raman scattering of pump photons.  
The most efficient cooling to the $|+\rangle$ state takes place  for $\Delta_\mathrm{c} = -\Omega_\mathrm{R}$ where the anti-Stokes photons are on resonance with the cavity.  At this point the net cooling and heating rates are,
\begin{eqnarray}
\Gamma_- =\frac{ 4 \chi^2 \nbar}{\kappa}+\frac{1}{2 T_2},\quad
\Gamma_+ =\frac{ \kappa \chi^2 \nbar}{(2 \Omega_R)^2+(\kappa/2)^2}+\frac{1}{2 T_2}, \label{eq:rates}
\end{eqnarray}
respectively, where $T_2=(1/2T_1+1/T_\varphi)^{-1}= 10.6 \ \mu$s is the (lab frame) dephasing rate. 
In equilibrium with the cavity bath, the final qubit polarization is given by detailed balance.

The state of the qubit was probed by measuring the phase shift of a microwave tone reflecting off of the cavity at the cavity resonance frequency.  The reflected signal was amplified by a lumped-element Josephson parametric amplifier (LJPA) operating in phase sensitive mode which allowed high fidelity, multi-state, single shot readout of the qubit state.  Our qubit sample exhibited excited (and higher excited) state population in excess of what would be expected from the nominal $T = 20$ mK environment.  For our measurements, we used post selection to disregard the higher excited state populations which was as much as 12\% of the qubit population.



\begin{figure} [htp]
\includegraphics[angle = 0, width = 0.4\textwidth]{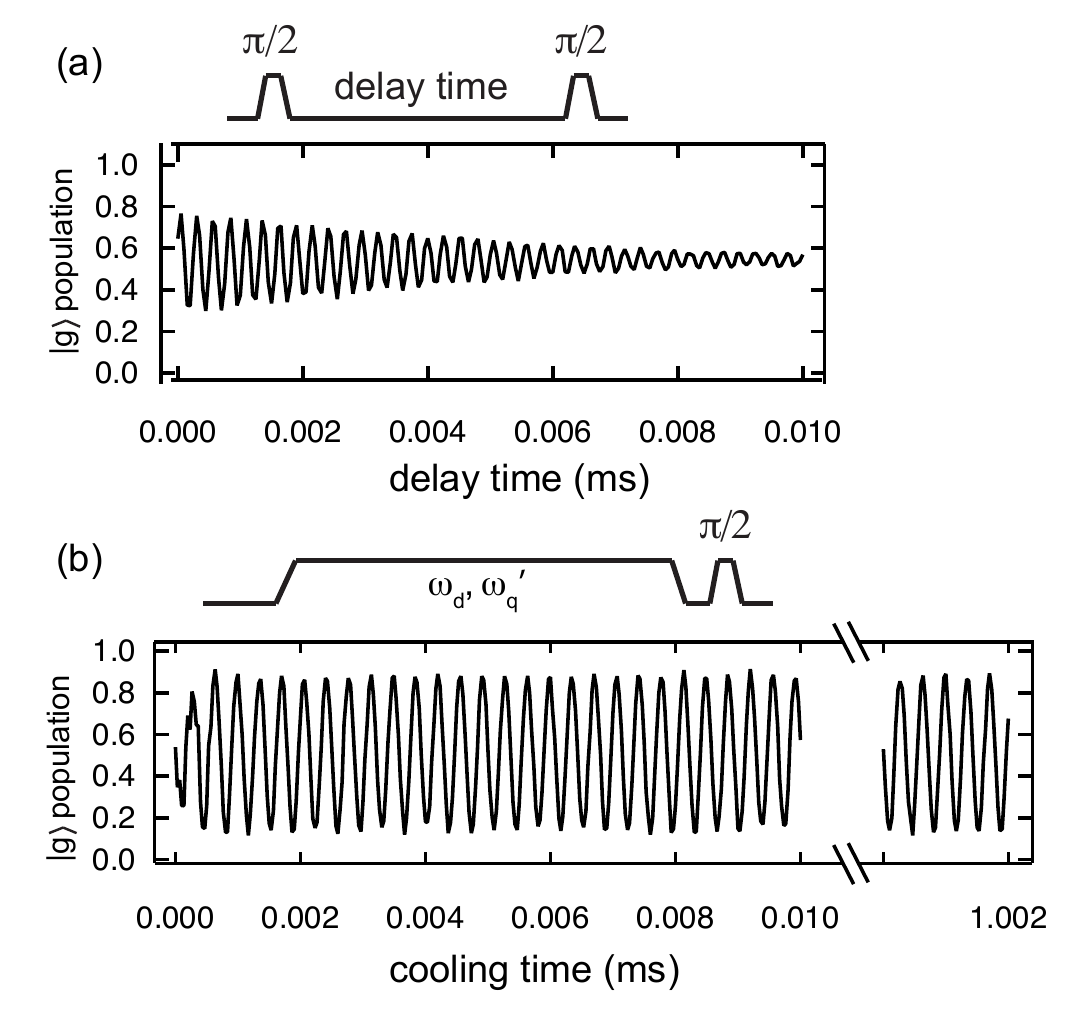}
\caption{\label{fig1r}   (a) Ramsey measurement using detuned pulses.  (b) Cavity cooling to the $|+\rangle$ state with $-\Delta_\mathrm{c}/2\pi = \Omega_\mathrm{R}/2\pi =9$ MHz, and $\nbar = 3.6$.  After driving the system at $\omega_\mathrm{q}'$ and $\omega_\mathrm{d}$ for a variable time, a detuned $\pi/2$ pulse  transfers the remaining coherence to the $\sigma_z$ basis.  High contrast, persistent ``Ramsey" fringes indicate that the qubit has been cooled to the $|+\rangle$ state.}
\end{figure}

 To demonstrate effective quantum bath engineering, we compare a Ramsey measurement 
  (Fig.\ 2(a)), to an experiment in which the qubit was cooled to the $|+\rangle$ state.  The Ramsey measurement consisted of two $\pi/2$ pulses detuned by $2.8$ MHz from the qubit frequency, followed by state readout in the $\sigma_z$ basis and showed a typical $T_2^* = 4.9\ \mu$s  exponential decay of coherence.  In Figure 2(b), the cavity was driven to cool the qubit to the $|+\rangle$ state.  After cooling the state for a variable period of time, the remaining coherence was measured by applying a $\pi/2$ pulse at a frequency detuned by $2.8$ MHz from $\omega_\mathrm{q}'$ and measuring the amplitude of the resulting oscillations in the ground state population.   We note that there is an initial build up of the coherence over a time scale of less than 1 $\mu$s given by $\Gamma^{-1} = (\Gamma_+ +\Gamma_-)^{-1}$ after which the system enters a steady state and the coherent oscillations persist indefinitely.


In Figure 3 we present tomography of the qubit state after it has come to equilibrium with the dissipative environment presented by the cavity.  The qubit was driven at a variable detuning $\Delta_\mathrm{q} = \omega_\mathrm{q}-\omega_\mathrm{r}$, where $\omega_\mathrm{r}$ is the drive frequency, and at variable cavity drive power detuned from the cavity by $\Delta_\mathrm{c}/2\pi  = -9$ MHz. The amplitude of the qubit drive was fixed to give $\Omega_\mathrm{R}/2\pi =+ 9$ MHz on resonance.    In Figure 3(a) we display the tomography data for $\langle \sigma_x\rangle$ and $\langle \sigma_z\rangle$  (note that in the rotating frame $\langle \sigma_y\rangle = 0$).  The dashed line indicates the dependence of $\omega_\mathrm{q}-\omega_\mathrm{q}'$ 
on the drive power, $P_\mathrm{d}$, indicating the detuning where the drive maintains resonance with the light-shifted qubit.  Following this curve, we plot $\langle \sigma_x\rangle$ and $\langle \sigma_z\rangle$ as a function of cavity photon number in Figure 3(b).  When $\nbar  = 0$, $\Gamma_+ = \Gamma_- = 1/(2 T_2)$ and the system is completely incoherent.  As the number of intracavity photons increases, coherence builds up along the $x$ direction and saturates around $\nbar=1$.  The 
 purity of the cooled state is given by $\Gamma_-/(\Gamma_++\Gamma_-)$. 
The observed maximum state purity was $70\%$ and was limited by our state readout fidelity ($\sim90\%$) and population relaxation to the second excited state of the transmon. The latter reduced the measured state purity by up to an additional $\sim20\%$ depending on the time delay between the tomography pulses and the readout.  The combination of these two effects led to a 70-80\% reduction in the measured state purity. Taking into account these reductions, 
 our measurements are close to the predicted value, plotted as a dashed line in Figure 3(b).

 \begin{figure*}[htp]
\includegraphics[angle = 0, width = .9\textwidth]{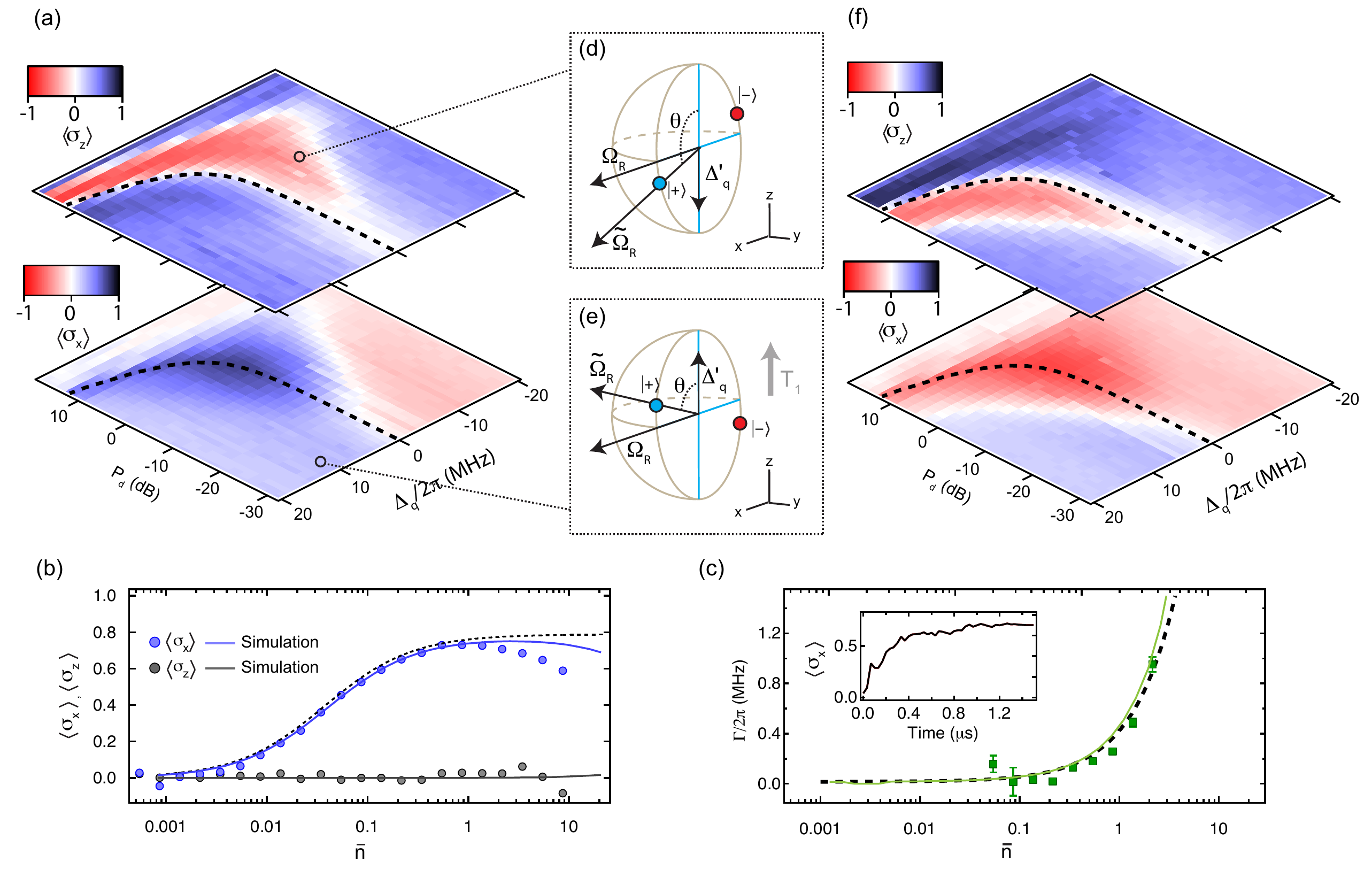}
\caption{\label{fig2} {\bf State tomography.}  (a) Color plots show $\langle \sigma_x\rangle$ and $\langle\sigma_z\rangle$ as a function of cavity drive power ($P_\mathrm{d} = 10 \log(\nbar)$ (dB) ) and qubit drive detuning for fixed cavity drive detuning, $-\Delta_\mathrm{c}/2\pi = \Omega_\mathrm{R}/2\pi = 9$ MHz.  The dashed line indicates the optimal detuning $\Delta_\mathrm{q}'=\omega_\mathrm{q}'-\omega_\mathrm{r}=0$. 
 (b,c)  $\langle \sigma_x\rangle$, $\langle\sigma_z \rangle$ and the cooling rate are plotted versus $\nbar$ for $\omega_\mathrm{r} = \omega_\mathrm{q}'$.  The dashed lines indicates the prediction from Eq.~(\ref{eq:rates}) which has been scaled by our measurement fidelity of 80\%.
    The solid blue, gray, and green lines indicate the results of the simulations for $\langle\sigma_x\rangle$, $\langle\sigma_z\rangle$ and the cooling rate respectively.  Error bars in (c) represent the estimated error in the exponential fit. (inset in c) $\langle\sigma_x\rangle$ vs.\  cooling time for $\nbar=1.4$. (d,e) Bloch sphere diagrams indicate that when the qubit drive is off resonance, the $|\pm\rangle$ states are tilted from the equator of the Bloch sphere.   (f) Color plot shows inversion of 
   $\langle \sigma_x\rangle$ and $\langle\sigma_z\rangle$ for $+\Delta_\mathrm{c}/2\pi = \Omega_\mathrm{R}/2\pi= 9$ MHz.}
\end{figure*}

When $|\chi\sqrt{\nbar}|>\kappa$, the system is in a regime of strong coupling where higher-order and rapidly-rotating terms that have so far been neglected become significant.  To explore this regime, we performed numerical simulations of the master equation.  These results are shown in Figure 3(b) as solid blue and gray lines and indicate that the maximum state purity is reduced compared to the predictions of Eq.~(2) at high drive powers.  Based on the simulation, we estimate that the \emph{actual} state purity was $94\%$, which corresponds to an effective temperature in the rotating frame of $T_\mathrm{eff}  = 150\ \mu\mathrm{K}$ for the dressed state.

To quantify the strength of the cavity damping, we plot the measured cooling rate in Figure 3(c), obtained by measuring the exponential timescale for buildup of ensemble population in the $|+\rangle$ 
state as the duration of the cooling pulse was increased (Fig.~3(c), inset).   The measured rate is in quantitative agreement with Eq.~(\ref{eq:rates}), shown as a dashed line, as long as $| \chi\sqrt{\nbar}|< \kappa$. At higher photon numbers, the observed increase in coherence was not exponential. In this regime, the system is expected to exhibit damped oscillations between $|-\rangle$ and $|+\rangle$ in analogy with vacuum Rabi oscillations (see the supplementary information).

When the qubit drive is off-resonant, the engineered dissipation drives the qubit to different points on the Bloch sphere.
As we illustrate in Figure 3(d), 
the off-resonant qubit drive creates an effective magnetic field at an angle $\theta = \arctan(\Omega_\mathrm{R}/\Delta'_\mathrm{q})$ with respect to the $z$ axis, tilting the $|\pm\rangle$ states from the equator of the Bloch sphere.  Here, $\Delta_\mathrm{q}' = \omega_\mathrm{q}'- \omega_\mathrm{r}$ is the detuning of the AC stark shifted qubit  frequency from the qubit drive.  In this case the cavity dissipation drives the system to the state obeying
%
%
$\sigma_\theta|+\rangle=(+1)|+\rangle$ where $\sigma_\theta \equiv \sin\theta\sigma_x + \cos\theta\sigma_z$.
%
%
%
When the cavity drive is very weak (Fig.~3(e))  
 dissipation due to the finite qubit $T_1 = 10\ \mu$s favors the $|-\rangle$ 
 state when $\Delta_\mathrm{q}'>0$.
 In Figure 3(f)   
  we display qubit state tomography when the cavity drive detuning, $+\Delta_\mathrm{c}/2\pi = \Omega_\mathrm{R}/2\pi = 9$ MHz.  In this case  the  cavity dissipation inverts the qubit to the $|-\rangle$ state.


    \begin{figure}
\includegraphics[angle = 0, width = .5\textwidth]{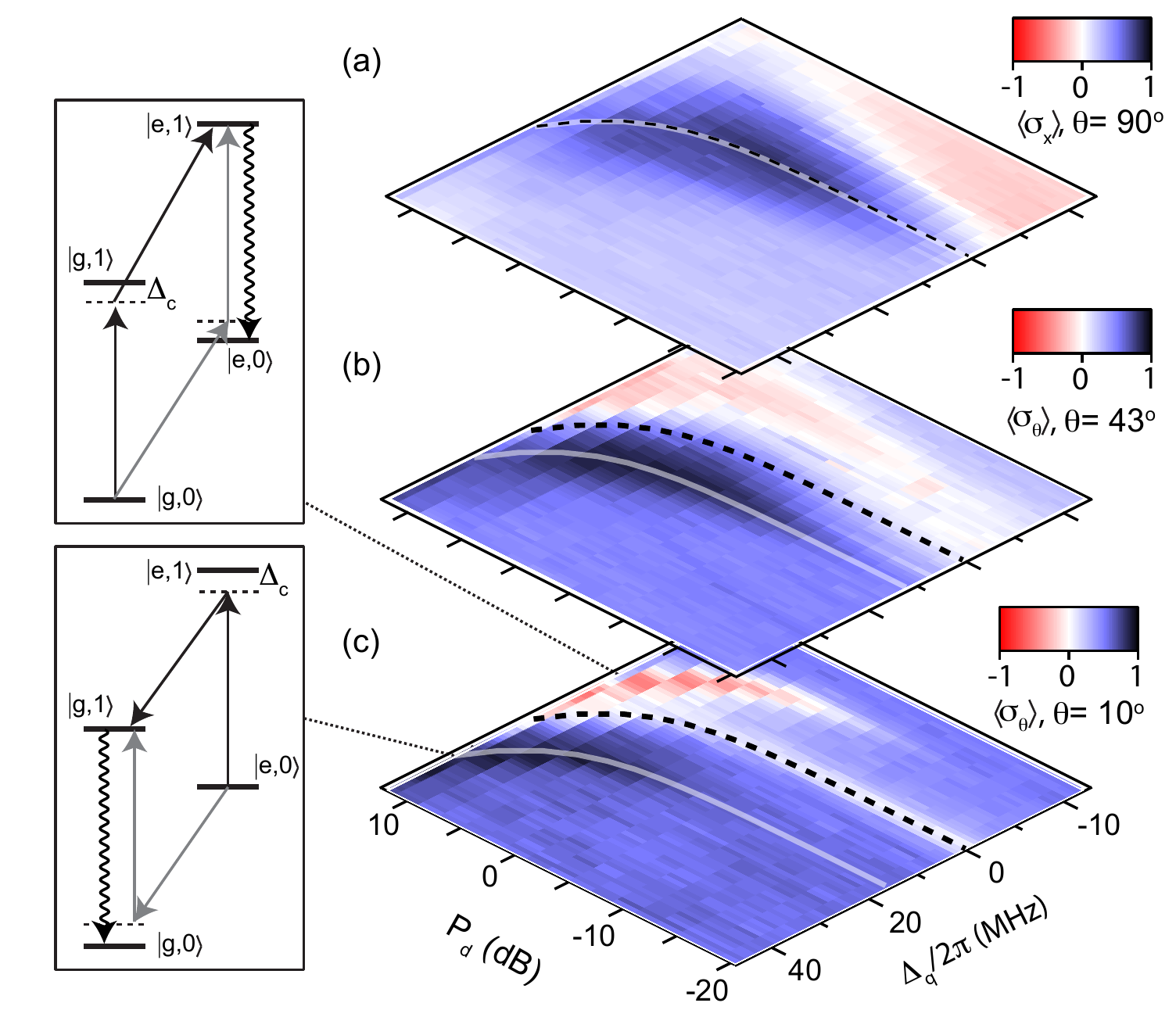}
\caption{\label{fig3} {\bf Preparing arbitrary superposition states using cavity dissipation.} Measurements of $\langle \sigma_\theta\rangle$ where $\sigma_\theta \equiv \sin\theta\sigma_x + \cos\theta\sigma_z$  
 for $\theta = \{90^\circ, 43^\circ,10^\circ\}$, (a-c), versus drive power and qubit-drive detuning.  The dashed lines indicate $\omega_\mathrm{q}'-\omega_\mathrm{q}$.
   The transparent gray lines indicate the qubit drive detuning that gives the most efficient cooling, $(\Delta_\mathrm{c}^2 - \Omega_\mathrm{R}^2)^{1/2}$.  For $\theta = 10^\circ$, two sideband transitions are visible corresponding to Raman transitions that leave the qubit in the ground (lower inset) or excited state (upper inset). The ground state Raman process (lower inset) involves absorption of a cavity drive photon, stimulated emission at $\omega_\mathrm{q}'  + \Delta_\mathrm{c}$, and emission into the cavity at $\omega_\mathrm{c}$. Two possible processes are labeled by gray and black arrows.}
\end{figure}

By driving the qubit off-resonance and altering the cavity detuning to remain equal to the off-resonant Rabi frequency, $\tilde{\Omega}_\mathrm{R} = \sqrt{\Omega_\mathrm{R}^2+\Delta_\mathrm{q}'^2}$, arbitrary superposition states of $|g\rangle$ and $|e\rangle$ can be prepared using the cavity dissipation.  As we show in the supplemental information, the heating and cooling rates are reduced by $(\Omega_\mathrm{R}/\Delta_\mathrm{q}')^2$ since the qubit drive is no longer resonant.
  In Figure 4 we display measurements that demonstrate cooling to arbitrary latitudes on the Bloch sphere.      Figure 4 displays $\langle \sigma_\theta \rangle$ for $\theta = \{90^\circ, 43^\circ, 10^\circ\}$ 
  for $\Delta_\mathrm{c}/2\pi=- 15$ MHz and variable drive power and detuning.  For $\theta = \pi/2$ the cooling is along $x$ as in Figure 3(a).  As $\theta$ is decreased (Fig.~4(b-c)),  the optimum cooling occurs for $\Delta_\mathrm{q}' = \sqrt{\Delta_\mathrm{c}^2 - \Omega_\mathrm{R}^2}$.

 For a weak Rabi drive strongly detuned from $\omega_\mathrm{q}'$, $\theta$ approaches zero, $|+\rangle \simeq |g\rangle$, 
 and the cavity dissipation mechanism crosses over to ordinary cavity-assisted Raman sideband cooling transitions similar to the transitions used to cool atoms\cite{wine79lasercool,neuh78sideband,died89zeropoint,hama98lattice} and superconducting qubits\cite{vale06sideband} using a strong atomic transition to enhance emission at a specific frequency.  Here, the cavity takes the place of the strong atomic transition.  In Figure 4(c), with $\theta = 10^\circ$, 
 two ``sideband" transitions emerge for $P_\mathrm{d} >0$ dB.  Level diagrams of the transitions are shown as insets in Figure 4 indicating that simultaneously detuning the drives from the cavity and qubit  allows selective optical pumping from the ground  or excited states. The rates for these transitions can be calculated using Fermi's golden rule and agree with a calculation based on Redfield theory (see the supplementary information).


As we previously noted, when $\Delta_\mathrm{c} = 0$, photon number fluctuations induce dephasing of the qubit in accordance with the theory of quantum measurement\cite{cler10noise}.  In this regime, cavity  photons convey information about the qubit state encoded as a phase shift corresponding to an elastic scattering event.  When the drive is detuned, the measurement is replaced by an inelastic scattering process in which the scattering of a photon into the cavity heralds a transition to the effective ground state of the system. 

In conclusion, we have demonstrated quantum bath engineering with a model two-level system.  The technique allows arbitrary superposition states of the the system to be prepared simply with saturating pulses.  In contrast to measurement based feedback\cite{wang01statestabilization,hofm98feedback,koro01feedback,sayr11feedback}, the technique is a form of coherent quantum feedback\cite{lloy00,mabu08} and is not limited by the quantum measurement efficiency.  State preparation fidelities in excess of 99.9\% are in principle possible with currently achievable sample parameters ($T_2 = 150\ \mu$s, $\kappa/2\pi =2$ MHz, $\Omega_\mathrm{R}/2\pi = 50$ MHz, $\chi/2\pi = -1$ MHz). Future multi-qubit implementations could enable the preparation of entangled many-body states suitable for quantum simulation and computation.

We thank R. Vijay for contributions to the LJPA and helpful comments. This research was supported in part by the National Science Foundation (DMR-1004406),
the U.S. Army Research Office (W911NF-11-1-0029 and W911NF-09-1-0514)
and the Office of the Director of National Intelligence
(ODNI), Intelligence Advanced Research Projects Activity
(IARPA), through the Army Research Office. All statements of fact,
opinion or conclusions contained herein are those of the
authors and should not be construed as representing the
official views or policies of IARPA, the ODNI, or the US
Government.


\end{document}


\title{Supplementary information for ``Cavity-assisted quantum bath engineering''} 

\author{K. W. Murch}
\affiliation{Quantum Nanoelectronics Laboratory, Department of Physics, University of California, Berkeley CA 94720}

\author{U. Vool}
\affiliation{Department of  Physics, Yale University, New Haven, CT 06520-8120}
\author{D. Zhou}
\affiliation{Department of  Physics, Yale University, New Haven, CT 06520-8120}
\author{S. J. Weber}
\affiliation{Quantum Nanoelectronics Laboratory, Department of Physics, University of California, Berkeley CA 94720}
\author{S. M. Girvin}
\affiliation{Department of  Physics, Yale University, New Haven, CT 06520-8120}
\author{I. Siddiqi}
\affiliation{Quantum Nanoelectronics Laboratory, Department of Physics, University of California, Berkeley CA 94720}

\date{\today}


\maketitle

\section{Sample fabrication and parameters}

The qubit sample was fabricated from aluminum deposited on high resistivity silicon in a two-step, double-angle evaporation process with an intervening oxidation step. The qubit consisted of a single Josephson junction with critical current $I_0 = 46$ nA shunting two paddles that set the charging energy ($E_\mathrm{C}/h = 145$ MHz) of the qubit and provide coupling to the cavity.  The coupling rate of the qubit to the cavity, $g = 70$ MHz was set by placing the sample away from the cavity electric field antinode.

The cavity was machined from aluminum (alloy 6061) and had dimensions of $35.6 \times 5.2\times28.1$ mm$^3$.  The cavity decay rate, $\kappa/2\pi = 4.3$ MHz, was set by adjusting the length of the center pin of a SMA coaxial connector protruding into the cavity volume.  The cavity was also addressed by a weakly coupled port which was used for the qubit manipulation pulses.

The dispersive coupling rate $\chi/2\pi = - 0.55 (.11)$ MHz was determined by measuring both the AC stark shift and measurement induced dephasing rate of a drive tone at the cavity resonance frequency as in our previous work\cite{vija12feedback}. We use $\chi/2\pi = -0.66$ MHz for comparison to theory and simulation because this value gave the best quantitative agreement.  

\section{Experiment setup and state measurement}
Drive signals were combined at room temperature and were sent to the sample via heavily attenuated coaxial lines.  The reflected signal from the strongly coupled cavity port was amplified by a near quantum limited lumped-element Josephson parametric amplifier (LJPA) operating in phase sensitive mode.  The LJPA was separated from the cavity by 4 cryogenic circulators and isolated from the HEMT amplifier at 2.7 K by three more isolators.   The LJPA was biased with a tone at 6.823 GHz and exhibited a gain of 19 dB with a 3 dB (instantaneous) 
 bandwidth of 30 MHz.  The phase of the LJPA pump was adjusted to amplify the quadrature of the reflected signal that contained information about the qubit states.

State measurement was performed by pulsing the readout and LJPA pump tones on.  After allowing 200 ns for transients to decay, a 200 ns section of data was integrated.  After repeating the experiment many times, a histogram of the readout voltage revealed three well-separated distributions corresponding to the qubit ground, excited, and higher excited transmon states. From these measurements we determined that the equilibrium population of the qubit states was 77\%, 14\%, and  9\% for the ground, excited and higher excited states respectively.  During cooling, the population of the higher excited states increased to $12\%$, owing to the the increase in the average excited state population.  In the experiment we use post-selection to remove the higher excited state population from our tomography measurements, but we note that coupling between the excited states results in a significant reduction of the measurement contrast.  We measured a $2 \ \mu$s timescale for equilibration between the excited states of the transmon, which results in a $\sim20\%$ reduction in the state purity determined by tomography.

We note that it is in principle possible to reduce the second excited state population using sideband cooling as demonstrated in Figure 3d simultaneously with cavity cooling to prepare pure states without the need to post select away population in higher qubit states.

In a typical experimental sequence the qubit and cavity drives were suddenly turned on for a 50 $\mu$s duration, followed by state tomography and readout.  The experiment was repeated  at 6 kHz.  Each data point in the tomography plots was the result of $10^5$ experimental sequences.

\section{Derivation of the system Hamiltonian \& Master Equation}
We begin our treatment with the Jaynes-Cummings Hamiltonian and two external
drives on the cavity, one with amplitude $\epsilon_\mathrm{d}$ and frequency $\omega_\mathrm{d}$ near the cavity resonance frequency $\omega_\mathrm{c}$,
and the other with amplitude $\epsilon_\mathrm{r}$ and frequency $\omega_\mathrm{r}$ near  the frequency of the qubit $\omega_\mathrm{q}$:
\begin{equation}
H=\omega_\mathrm{c}a^{\dagger}a-\frac{\omega_\mathrm{q}}{2}\sigma_{z}+g(a^{\dagger}\sigma_{-}+a\sigma_{+})+\left[\epsilon_\mathrm{d}e^{-i\omega_\mathrm{d}t}a^{\dagger}
+\epsilon_\mathrm{r}e^{-i\omega_\mathrm{r}t}a^{\dagger}+\mathrm{h.c}\right] ,
\end{equation}
where $\sigma_{+}$ is the operator exciting the qubit and $\sigma_{-}$ de-exciting it. 
By including qubit and cavity decay terms we can write the master equation\cite{blai07proces}:
\begin{equation}
\dot{\rho}=-i\left[H,\rho\right]+\kappa\mathcal{D}[a]\rho+\frac{\Gamma_{\varphi}}{2}\mathcal{D}[\sigma_{z}]\rho
+\Gamma_{-}\mathcal{D}[\sigma_{-}]\rho +\Gamma_{+}\mathcal{D}[\sigma_{+}]\rho,
\end{equation}
where $\mathcal{D}[L]\rho=(2L\rho L^{\dagger}-L^{\dagger}L\rho-\rho
L^{\dagger}L)/2$, $\Gamma_{\varphi}=\frac{1}{T_{\varphi}}$ is the phase relaxation rate and the population relaxation rate is given by $\frac{1}{T_{1}}=\Gamma_{-}+\Gamma_{+}$.

 By applying the dispersive shift,
$U=e^{\frac{g}{\Delta}(a\sigma_{+}-a^{\dagger}\sigma_{-})}$, where
$\Delta=\omega_\mathrm{q}-\omega_\mathrm{c}$, and moving to the rotating frames for
both the qubit and cavity, $U_\mathrm{c}=e^{ia^{\dagger}a\omega_\mathrm{d}t},\; U_\mathrm{q}=e^{-i\sigma_{z}\frac{\omega_\mathrm{r}}{2}t}$, we can simplify the Hamiltonian.
Taking terms only up to second order in $\frac{g}{\Delta}$ we are left with
(assuming $\epsilon_\mathrm{r}$ and $\epsilon_\mathrm{d}$ are real):
\begin{equation}
H=-\Delta_\mathrm{c}a^{\dagger}a-\frac{\Delta_\mathrm{q}+\chi}{2}\sigma_{z}-\frac{\Omega_\mathrm{R}}{2}\sigma_{x}-\chi a^{\dagger}a\sigma_{z}+\epsilon_\mathrm{d}(a^{\dagger}+a),
\end{equation}
where, $\Omega_\mathrm{R}=-\frac{2\epsilon_\mathrm{r}g}{\Delta}$, $\chi=\frac{g^2}{\Delta}$, $\Delta_\mathrm{c}=\omega_\mathrm{d}-\omega_\mathrm{c}$, $\Delta_\mathrm{q}=\omega_\mathrm{q}-\omega_\mathrm{r}$, and we have ignored terms rotating at $\omega_\mathrm{r}-\omega_\mathrm{d}$ or higher frequencies.
%
%
Note that the second-order correction also includes the well-known Purcell effect\cite{purc46effect} which adds a term proportional to $\kappa$ to the qubit relaxation rate due to  the coherent mixing of qubit and cavity excitations. 

We eliminate the drive term by displacing the field operator $a=\bar{a}+d$.  Note that this displacement acting on the cavity dissipator term introduces
\begin{equation}
\kappa \mathcal{D}[a]\rho = \kappa \mathcal{D}[d]\rho
+ \frac{\kappa}{2} \left[\left( \bar{a}^{\star} d - \bar{a} d^\dagger\right),\rho\right].
\end{equation}
Because the last term has the form of a commutator, it is equivalent to a shift in the Hamiltonian of
\begin{equation}
\Delta H = \frac{i\kappa}{2}\left({\bar a}^{\star}d - {\bar a}d^\dagger\right).
\end{equation}
 A convenient displacement choice will be:
\begin{equation}
\bar{a} = \frac{\epsilon_\mathrm{d}}{\Delta_\mathrm{c}+i\kappa/2}.
\end{equation}
With this choice we eliminate the drive and all of the terms linear in $d$ in $H+\Delta H$ except for
 terms of the form $d\sigma_{z}$. Our
Hamiltonian can now be written as:
\begin{equation}
H=-\Delta_\mathrm{c}d^{\dagger}d-\frac{\Delta_\mathrm{q}+\chi(2\bar{n})}{2}\sigma_{z}-\frac{\Omega_\mathrm{R}}{2}\sigma_{x}-\chi(\bar{a}^{\star}d+\bar{a}d^{\dagger}+d^{\dagger}d)\sigma_{z},
\label{eq:Hamil}
\end{equation}
where $\bar{n}=|\bar{a}|^{2}$, and we have absorbed the Lamb shift into the definition of the qubit frequency [via the replacement $(2\bar n + 1) \rightarrow (2\bar n)$]. The Hamiltonian above is the one used in our simulations.

To better understand this Hamiltonian and present a more intuitive understanding of the cooling process we choose the Rabi drive frequency such that $\Delta_\mathrm{q}'=\Delta_\mathrm{q}+\chi(2\bar{n})=0$, 
meaning the qubit drive is exactly on resonance if there are $\bar{n}$ photons in the cavity.  By performing a Hadamard rotation to interchange $\sigma_{z}$ and $\sigma_{x}$ and writing the new $\sigma_{x}$ as $\sigma_{+}+\sigma_{-}$, we arrive at an effective Hamiltonian (ignoring rapidly rotating terms as we choose the cavity drive (red) detuning to match the Rabi frequency, $\Delta_\mathrm{c}=-\Omega_\mathrm{R}$),
\begin{equation}
H=-\Delta_\mathrm{c}d^{\dagger}d-\frac{\Omega_\mathrm{R}}{2}\sigma_{z}-\chi(\bar{a}^{\star}d\sigma_{+}+\bar{a}d^{\dagger}\sigma_{-}).   \label{eq:effectiveJC}
\end{equation}
The new Hamiltonian is an effective Jaynes-Cummings model in which the effective qubit ground state is the lower energy eigenstate of $\sigma_{z}$ [$\sigma_{x}$ in the original (rotating) spin frame before the Hadamard transformation].
With the Hamiltonian in this form, one understands that the cooling scheme is simply the dissipation of the dressed qubit to its effective ground state via photon emission into the cavity, just as in the ordinary (un-driven) Jaynes-Cummings model.  [Here however the photon emission is actually Raman scattering of the cavity pump photons.] Using the fact that the peak value of the resonator density of states is $\rho=\frac{2}{\pi\kappa}$ Fermi's Golden Rule yields for this particular resonant case, the following simple expression for the cooling rate
\begin{equation}
\Gamma = 2\pi |\chi\bar{a}|^2 \rho = \frac{4\chi^2}{\kappa}\bar{n},
\end{equation}
which is valid in the limit of weak coupling ($|\chi|\sqrt{\bar n}\ll\kappa)$.

\section{General Heating and cooling rates}
Before extending the above results to the more general case of arbitrary detunings for the cavity and qubit drives, it is convenient to reformulate the above derivation.
Without making the Hadamard transformation and the cavity displacement transformation mentioned above, the Hamiltonian for the qubit alone can be written in the frame rotating at the Stark-shifted qubit frequency as: 
\begin{align}
H_\mathrm{q} = -\frac{\Omega_\mathrm{R}}{2}\sigma_x - \frac{2\chi [a^\dagger a-\bar n]}{2} \sigma_z,   
\end{align}
in which the dispersive coupling term can be thought of as adding photon shot noise that causes dephasing of the qubit \cite{cler10noise,blai04archit}.
Following the derivation of Fermi's Golden Rule in terms of noise spectral densities presented in\cite{cler10noise},
the qubit excitation and de-excitation rates can be calculated from the spectral density of the noise perturbing the qubit in directions orthogonal to the $x$ quantization direction.
Spectral density at negative (positive) frequencies correspond\cite{cler10noise} to energy emitted (absorbed) by the bath: 
\begin{align}
\Gamma_\pm &= \frac{1}{4}\left\{\tilde S_{zz}(\mp\Omega_\mathrm{R}) +\tilde S_{yy}(\mp\Omega_\mathrm{R})\right\}, \\
    &= \frac{1}{4}\left\{4\chi^2{S}_{nn}(\mp\Omega_\mathrm{R}) +
    \tilde{S}^{'}_{zz}(\mp\Omega_\mathrm{R}) + \tilde S_{yy}(\mp\Omega_\mathrm{R})\right\},
\end{align}
where $\tilde S_{zz}$ is the noise power spectral density related to the
autocorrelation function of the noisy coefficient of $\sigma_z$ in the Hamiltonian.
It can be broken into two parts: one due to the photon number
fluctuations, i.e., $S_{nn}$ and the other one due to all
other processes $\tilde{S}^{'}_{zz}$. 
$\Gamma_-$ refers to the transition rate from the high-energy eigenstate of $\sigma_x$ with
$-1$ eigenvalue to the low-energy state with $+1$ eigenvalue. 
This cooling transition is the dominant one for our assumption of a red-detuned cavity drive, i.e.,
$\Delta_\mathrm{c} = -\Omega_\mathrm{R}$.

Being cognizant\cite{ithi05decoher} that $S_{yy}$ is evaluated in the rotating frame and with further approximations that:  
\begin{align}
&\tilde{S}_{yy}(\mp\Omega_\mathrm{R}) \simeq \tilde S_{yy}(0)= \Gamma_1 ,\\ 
&\tilde{S}^{'}_{zz}(\mp\Omega_\mathrm{R}) \simeq \tilde{S}^{'}_{zz}(0)=2\Gamma_\varphi ,\\ 
&S_{nn}[\omega] =  \frac{{\bar n}\cdot\kappa}{(\kappa/2)^2+(\omega+\Delta_\mathrm{c})^2} ,
\end{align}
where $\Gamma_\varphi$ is the pure-dephasing rate due to other sources, and we take $S_{nn}$ to be the result for an uncoupled driven cavity. 
Note that noise along the $z$ direction is not affected by the transformation to the rotating frame. 
Using these terms we obtain\cite{ithi05decoher}:
\begin{align}
\Gamma_- = & \frac{4\chi^2}{\kappa}{\bar n} + \frac{\Gamma_\varphi}{2} +
                    \frac{\Gamma_1}{4}, \\
\Gamma_+ = & \frac{\kappa\chi^2}{(\kappa/2)^2+4\Omega_\mathrm{R}^2}{\bar n} +
                \frac{\Gamma_\varphi}{2} + \frac{\Gamma_1}{4} .
\end{align}
For the cooling to be effective, we need:
\begin{align}
\frac{4\chi^2}{\kappa}{\bar n} \gg \frac{\Gamma_\varphi}{2} +
                    \frac{\Gamma_1}{4} . \label{eq:rabi_cooling}
\end{align}
i.e., the shot-noise term in $\Gamma_-$ should dominate so that the asymmetry in $S_{nn}(\pm\Omega_\mathrm{R})$ strongly affects the transition rates.  

If the qubit is driven off-resonance by a detuning $\Delta_\mathrm{q}'$ from its (Stark-shifted) frequency,  the Hamiltonian takes the form:
\begin{align}
H_\mathrm{q} = -\frac{\Delta_\mathrm{q}'}{2}\sigma_z- \frac{\Omega_\mathrm{R}}{2}\sigma_x
        - \frac{2\chi [a^\dagger a-\nbar]}{2} \sigma_z  .
\end{align}
If the cavity detuning is set to the new Rabi frequency $\Delta_\mathrm{c} = \pm\sqrt{\Omega_\mathrm{R}^2+\Delta_\mathrm{q}'^2}$, it is possible to cool the qubit to
an arbitrary position on the Bloch sphere using the same mechanism.  The cooling and heating will occur between the two states $\ket \pm$, which are the
eigenstates of $(\Delta_\mathrm{q}'\sigma_z + \Omega_\mathrm{R}\sigma_x)/\sqrt{\Omega_\mathrm{R}^2+\Delta_\mathrm{q}'^2}$
with eigenvalue $\pm1$.\\
The transition rates can be calculated in a manner identical to the on-resonance case and are found to be:
\begin{align}
\Gamma_\pm = &\frac{1}{4}\left\{
        \tilde{S}_{zz}\left(\mp\sqrt{\Omega_\mathrm{R}^2+\Delta_\mathrm{q}'^2}\right)\sin^2\theta
        +\tilde S_{xx}\left(\mp\sqrt{\Omega_\mathrm{R}^2+\Delta_\mathrm{q}'^2}\right)\cos^2\theta
        + \tilde S_{yy}\left(\mp\sqrt{\Omega_\mathrm{R}^2+\Delta_\mathrm{q}'^2}\right)
            \right\}, \\
       \simeq& \chi^2S_{nn}(\mp\Omega_\mathrm{R}/\sin\theta)\sin^2\theta
                 + \frac{1}{2}\Gamma_\varphi\sin^2\theta
                 + \frac{1}{4}\Gamma_1\left(1+\cos^2\theta\right), \label{eq:g_pm}
\end{align}
where we define $\tan\theta = \Omega_\mathrm{R}/\Delta_\mathrm{q}'$.

With weak qubit drive at large detuning $\Delta_\mathrm{q}'\gg\Omega_\mathrm{R}$, the qubit is no longer dressed and cooling is along the $z$
direction.  In this limit, the physics devolves into the ordinary sideband cooling process in which both the qubit and cavity   
drives are detuned by $\sim\Delta_\mathrm{q}'$.
The rates are given by:
\begin{align}
\Gamma_- &\simeq \frac{4\chi^2{\bar n}}{\kappa}
            \left(\frac{\Omega_\mathrm{R}}{\Delta_\mathrm{q}'}\right)^2
       + \frac{\Gamma_\varphi}{2}\left(\frac{\Omega_\mathrm{R}}{\Delta_\mathrm{q}'}\right)^2
            + \frac{\Gamma_1}{2}, \label{eq:g+}\\
\Gamma_+ &\simeq\frac{\kappa\chi^2{\bar n}}{(\kappa/2)^2+4\Delta_\mathrm{q}'^2}
        \left(\frac{\Omega_\mathrm{R}}{\Delta_\mathrm{q}'}\right)^2
       + \frac{\Gamma_\varphi}{2}\left(\frac{\Omega_\mathrm{R}}{\Delta_\mathrm{q}'}\right)^2
            + \frac{\Gamma_1}{2} \label{eq:g-}.
\end{align}
Note that the cooling is less effective than the resonant case, as we have an additional factor of $\left(\frac{\Omega_\mathrm{R}}{\Delta_\mathrm{q}'}\right)^2$ which suppresses the rate.

\section{Raman transition rates}
To complete the picture, we provide a derivation of the transition rates for Raman sideband cooling to show that they indeed coincide with
the Rabi cooling for $\Delta_\mathrm{q}'\gg\Omega_\mathrm{R}$.
The Hamiltonian can be written as:
\begin{align}
H =& H_0 + V, \\
H_0    =& -\frac{\Delta_\mathrm{q}'}{2}\sigma_z - \Delta_\mathrm{c} 
 d^\dagger d
            - \frac{\Omega_\mathrm{R}}{2} \sigma_x, \\
V =& -\chi \left({\bar a}d^\dagger+ {\bar a}^{\star}d + d^\dagger d\right)\sigma_z.
\end{align}
Here we have displaced the cavity field $a = {\bar a} + d$, with
${\bar a} = \epsilon_\mathrm{d}/(\Delta_\mathrm{c}+i\kappa/2)$. 
The cooling/heating processes rely on the fact that the dispersive shift term couples
the qubit and cavity.
We thus treat the dispersive term as a perturbation and use Fermi's golden rule to
calculate the transition rates.

The qubit part of $H_0$ can be further diagonalized and we get:
\begin{align}
{\ket{\tilde e}} \simeq \ket{e} + \frac{\Omega_\mathrm{R}}{2\Delta_\mathrm{q}' } \ket{g}, \\
{\ket{\tilde g}} \simeq  \ket{g} - \frac{\Omega_\mathrm{R}}{2\Delta_\mathrm{q}' } \ket{e},
\end{align}
where we have used the fact that $\Omega_\mathrm{R}\ll\Delta_\mathrm{q}$.  Note that,
\begin{align}
\left|\bra{\tilde{e},1}V\ket{\tilde{g},0}\right|^2 =
\left|\bra{\tilde{e},0}V\ket{\tilde{g},1}\right|^2 = \chi^2{\bar n}
            \left(\frac{\Omega_\mathrm{R}}{\Delta_\mathrm{q}' }\right)^2 .
\end{align}
We also make use of the effective cavity density of states $\rho(\omega)=-\frac{1}{\pi}\operatorname{Im}\frac{1}{\omega-\Delta_\mathrm{q}'+i\frac{\kappa}{2}}$.
Applying Fermi's golden rule, we get the same cooling/heating rates as in the noise
spectral density calculation, i.e.,
\begin{align}
\Gamma_- = & \frac{4\chi^2}{\kappa}{\bar n}
            \left(\frac{\Omega_\mathrm{R}}{\Delta_\mathrm{q}'}\right)^2, \\
\Gamma_+ = & \frac{\kappa\chi^2}{(\kappa/2)^2+4\Delta_\mathrm{q}'^2}{\bar n}
        \left(\frac{\Omega_\mathrm{R}}{\Delta_\mathrm{q}'}\right)^2.
\end{align}

\begin{figure*}
\includegraphics[angle = 0, width = .6\textwidth]{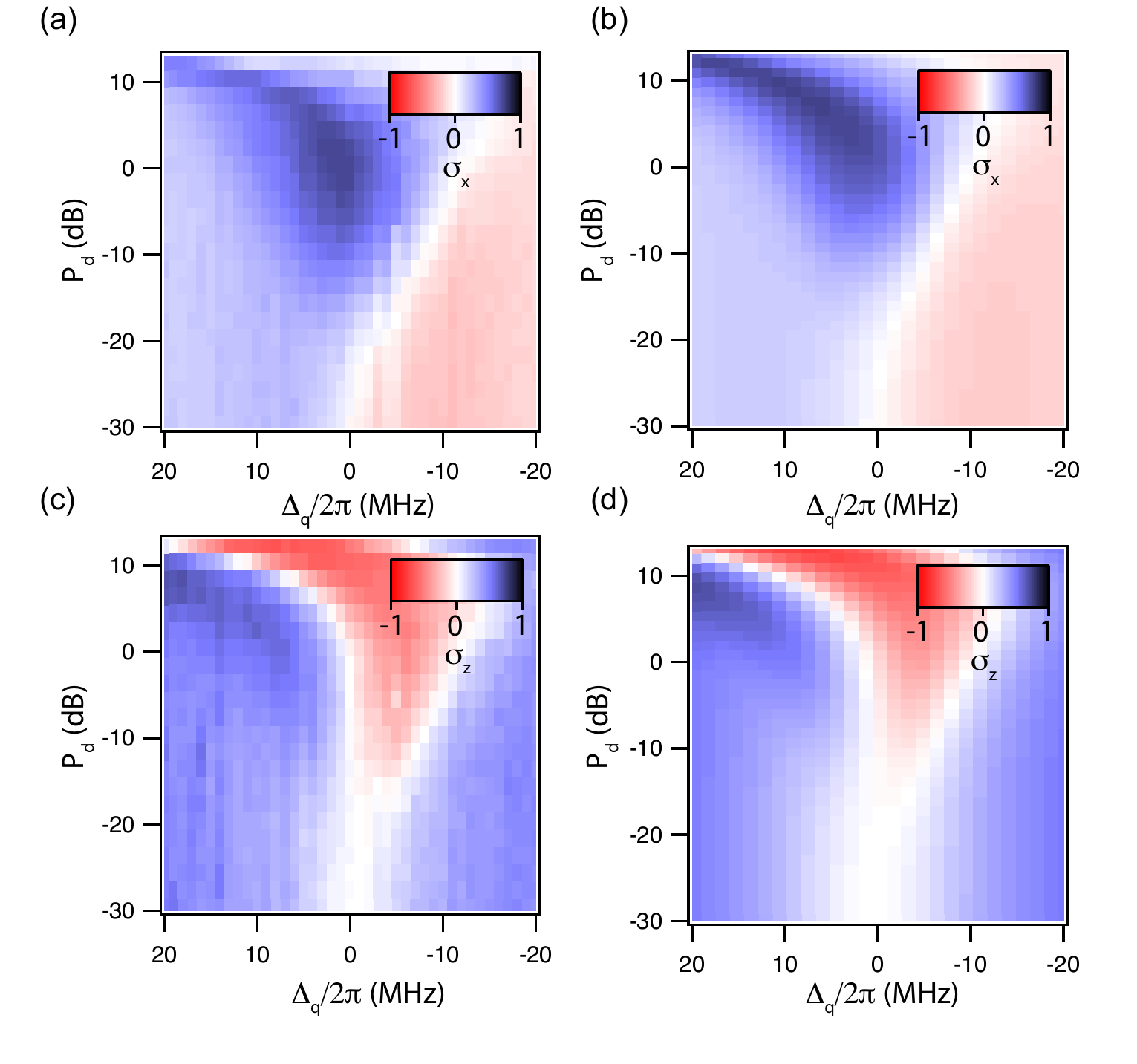}
\caption{\label{fig1} {\bf Comparison between theory and experiment}  (a,b) Color plots show $\langle \sigma_x\rangle$ as a function of cavity drive power and qubit drive detuning for fixed cavity drive detuning, $-\Delta_\mathrm{c} = \Omega_\mathrm{R} = 2\pi \times 9$ MHz for the experiment (a) and simulation (b). (c,d) Color plot of $\langle \sigma_z\rangle$ for experiment (c) and simulation (d).  Simulation results in (b) have been scaled by $0.8$ to account for imperfections in the tomography.}  
\end{figure*}

\section{Simulations of the master equation}
A set of numerical simulations were performed to better understand our experimental results. All the simulations involved numerically solving
the Master equation presented in Section III 
and especially the Hamiltonian in Eq.~(\ref{eq:Hamil}). The simulations were written in python and make significant use of the QuTip toolbox\cite{joha12qutip}.
Figure 1 shows a comparison between the tomography results produced by the simulation and the experiment for $\langle \sigma_x\rangle$ in Figure 1a,b
and $\langle \sigma_z\rangle$ in Figure 1c,d. The experimental results are seen to be consistent with the theoretical predictions of our model.

\begin{figure*}
\includegraphics[angle = 0, width = .6\textwidth]{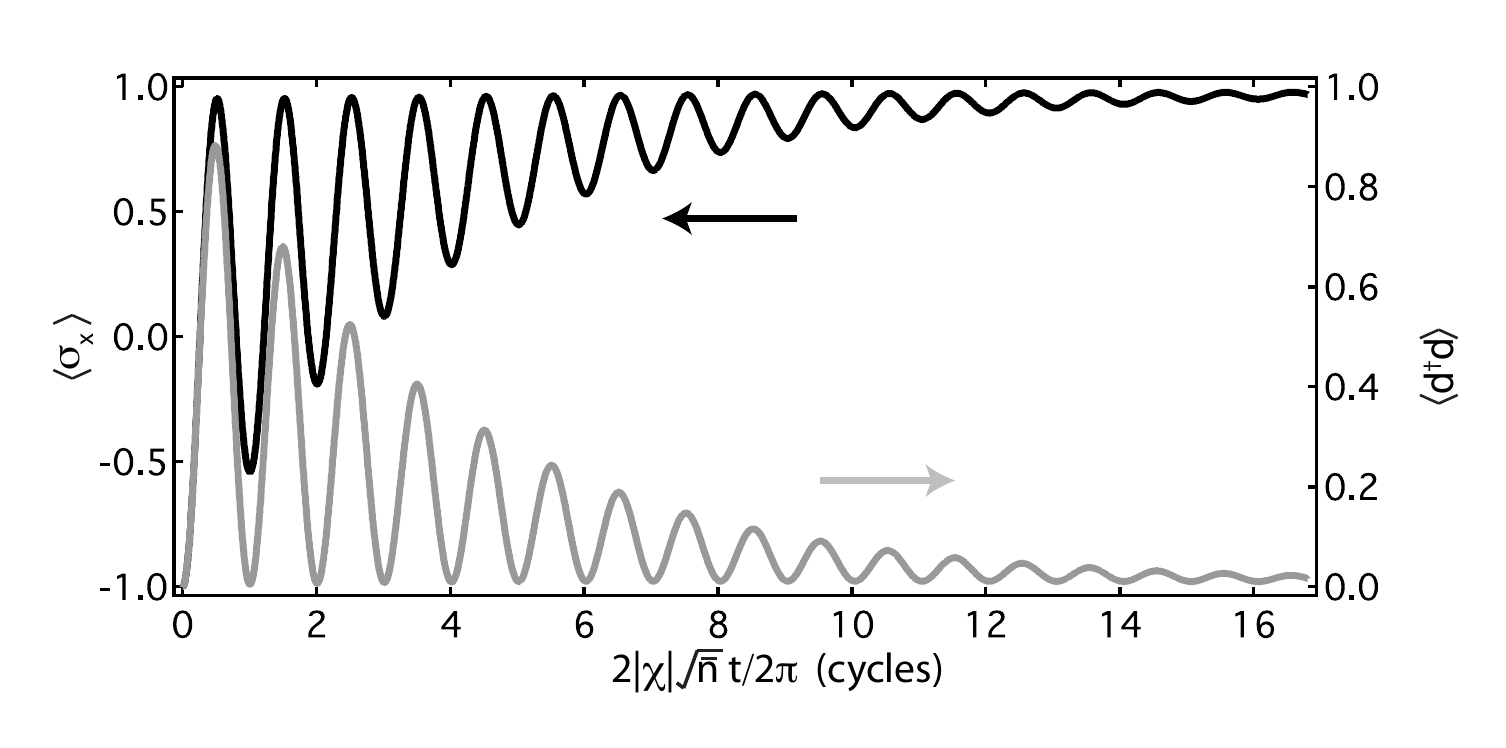}
\caption{\label{fig2} {\bf Strong coupling time domain oscillations}  The figure shows the simulated time domain evolution of $\langle \sigma_x\rangle$ (black) and $\langle d^{\dagger}d\rangle$ (grey). The total photon number in the cavity was taken to be $\nbar=3.31$ and $\kappa/2\pi = 0.2$ MHz. The oscillations between cavity photon and excited state qubit are clearly visible and the frequency of oscillation matches the theoretical prediction $2|\chi|\sqrt{\nbar}$, the energy splitting in the effective Jaynes-Cummings model.}
\end{figure*}

\section{Strong coupling regime}
The theoretical calculations in the sections above are all based on the weak coupling assumption $\sqrt{\nbar}|\chi|\ll\kappa$, namely that the coupling constant in the effective
Hamiltonian [e.g. in Eq.~(\ref{eq:effectiveJC})] 
is much smaller than the decay rate. When this assumption starts to fail the calculated decay rates are no longer a good model for the behavior of the system,
which instead of exponential decay to the effective ground state shows oscillations\cite{brun96vacuum} in the time domain. Intuitively this can be described
as a coherent oscillation of the system (following a sudden switch of the coupling) between two states.  The first state has the qubit in the high-energy eigenstate of $\sigma_{x}$ and no Raman photon in the cavity.  The second has the qubit in the low-energy dressed state and one Raman photon in the cavity.  The coupling is strong enough that the system coherently oscillates back and forth between these states multiple times before the cavity decay eventually
brings the system to equilibrium. This regime can be reached experimentally with currently available parameters (strong drives and a low $\kappa$ cavity).
Figure 2 shows a simulation of the system in the strong coupling regime in which this effect can be observed.  One sees excellent agreement of the oscillation rate with the predicted effective `vacuum Rabi frequency' $2|\chi|\sqrt{\nbar}$. 


\bibliographystyle{naturemag}
\bibliography{supplementaryref}